\documentclass[twocolumn,showpacs,preprintnumbers,amsmath,amssymb,prl]{revtex4-1}

\usepackage{graphicx}
\usepackage{dcolumn}
\usepackage{bm}
\usepackage{tabularx}
\usepackage{amsmath}

\begin{document}

\title{Bulk Fermi surface of charge-neutral excitations in SmB$_6$ or not:\\ a heat-transport study}

\author{Y. Xu,$^1$ S. Cui,$^1$ J. K. Dong,$^1$ D. Zhao,$^{2,3}$ T. Wu,$^{2,3,4}$ X. H. Chen,$^{2,3,4}$ Kai Sun,$^5$ Hong Yao,$^6$ and S. Y. Li$^{1,4,*}$}

\affiliation{$^1$State Key Laboratory of Surface Physics, Department of
Physics, and Laboratory of Advanced Materials, Fudan University,
Shanghai 200433, China \\
$^2$Hefei National Laboratory for Physical Science at Microscale and Department of Physics, University of Science and Technology of China,
Hefei, Anhui 230026, China\\
$^3$Key Laboratory of Strongly-coupled Quantum Matter Physics, University of Science and Technology of China, Chinese Academy of Sciences,
Hefei 230026, China\\
$^4$Collaborative Innovation Center of Advanced Microstructures, Nanjing University, Nanjing 210093, China\\
$^5$Department of Physics, University of Michigan, Ann-Arbor, Michigan 48109, USA\\
$^6$Institute for Advanced Study, Tsinghua University, Beijing 100084, China}

\date{\today}

\begin{abstract}
Recently there have been increasingly hot debates on whether a bulk Fermi surface of charge-neutral excitations exists in the topological Kondo insulator SmB$_6$. To unambiguously resolve this issue, we performed the low-temperature thermal conductivity measurements of a high-quality SmB$_6$ single crystal down to 0.1 K and up to 14.5 T. Our experiments show that the residual linear term of thermal conductivity at zero field is {\it zero}, within the experimental accuracy. Furthermore, the thermal conductivity is insensitive to magnetic field up to 14.5 T. These results exclude the existence of fermionic charge-neutral excitations in bulk SmB$_6$, such as scalar Majorana fermions or spinons, thus put a strong constraint on the explanation of the quantum oscillations observed in SmB$_6$.
\end{abstract}


\maketitle

Topological insulator is a novel quantum state of matter, and has been suggested theoretically and observed experimentally \cite{XLQi1,Hasan,Bernevig,Moore,Fu}.
Like the edge channel found in the quantum Hall system, the strong spin-orbit coupling in a three dimensional topological insulator leads to a non-trivial and robust conducting surface state. This metallic state is protected by the time-reversal symmetry. Studies on topological insulators have later stimulated the search for many other topological materials, such as topological crystalline insulators, Weyl and Dirac semimetals, and topological Kondo insulator \cite{Wanxg,Wenghm,MDzero,LFu}. Especially, interaction effect could play an important role in topological Kondo insulators and render exotic physics in them.

As one of the most historical heavy-fermion (HF) materials, SmB$_6$ has been studied for more than 50 years \cite{Jaccarino,Menth} and was recently shown to be a topological Kondo insulator \cite{MDzero-review}. For decades, the low-temperature conductivity in SmB$_6$ remains puzzling: its resistivity shows insulating behavior down to a few Kelvins but saturates down to the lowest temperature upon further reducing the temperature. This puzzle was successfully resolved by recent transport experiments, which show that the material is a bulk insulator but with a metallic surface \cite{Swolgast,DJKim2}, consistent with the theoretical prediction that SmB$_6$ is a topological Kondo insulator \cite{MDzero,MDzero2,VAlexandrov1,FLu}. At high temperatures, transport properties are dominated by thermal excitations in the insulating bulk, and thus insulating behaviors are observed. At low temperatures, however, bulk excitations vanishes because of the energy gap, and surface signals become dominant. The existence of the metallic surface states is now well established and observed in a number of experiments\cite{XHZhang, GLi,BSTan,FChen,JJiang, Neupane,NXuprb,EFrantzeskakis, Denlinger1,Denlinger2,STM,Sahana,STM2,thermalelectric}.

Although electric transport measurements so far show that the bulk of SmB$_6$ has no gapless charge carriers (namely a finite charge-gap), there exist other experiments suggesting possible gapless excitations in the bulk. Especially, the recent quantum oscillation measurement claims multiple Fermi seas in the bulk of an insulating SmB$_6$ sample \cite{BSTan}, in direct contradiction with transport measurements. As the electrical transport only measures the charge degrees of freedom and magnetization focuses more on spins, it is possible that the bulk of the sample may have some gapless and charge-neutral degrees of freedom. If these neutral modes form structures similar to a Fermi sea, they may produce de Haas-van Alphen (dHvA) oscillation signals. Baskaran \cite{GBaskaran} proposed that the bulk of SmB$_6$ may form a Majorana Fermi liquid or a spin liquid with spinon Fermi surfaces, which is a highly nontrivial quantum state of matter\cite{Maissam13}. However, such charge neutral modes have not yet been observed in any other measurements, such as neutron scattering \cite{neutron}. Their existence in SmB$_6$ is still under hot debate such that experimentally resolving this issue is highly desired.

To unambiguously solve this puzzle, we performed the low-temperature thermal conductivity study of a high-quality SmB$_6$ single crystal. In contrast to electrical transport, thermal conductivity probes both charge and charge-neutral degrees of freedom, and thus is an ideal approach to reveal {\it unbiased} information about low-energy excitations in the system. In particular, as discussed above, one key question in the study of SmB$_6$ is whether its bulk has nontrivial charge-neutral gapless degrees of freedom. As the thermal conductivity from these gapless modes is expected to show different scaling behaviors (e.g. linear in temperature $\sim T$ for charge-neutral fermionic excitations) in contrast to the $T^3$ behavior from phonon contributions, our measurement can directly probe and detect these neutral modes, if they exist in the system. Previously, a heat transport measurement on SmB$_6$ was done at and above 1.5 K \cite{kappaold}, and the thermal conductivity was found to be dominated by phonons. However, this temperature is too high to compare with the recent dHvA data, and thus cannot provide a conclusive answer for the puzzle in SmB$_6$.

In this Letter, we report the low-temperature thermal conductivity measurements of a high-quality SmB$_6$ single crystal down to 0.1 K. No residual linear term $\kappa_0/T$ is observed at zero magnetic field. Furthermore, the thermal conductivity is insensitive to magnetic field up to 14.5 T. The absence of $\kappa_0/T$ unambiguously demonstrates that no fermionic excitations exist in the bulk SmB$_6$, which we think can clearly settle down the debate. We shall discuss other possible explanations of the quantum oscillations observed in SmB$_6$ below.

The SmB$_6$ single crystal used in this work is from the same batch for previous dHvA and magnetoresistance measurements \cite{GLi,FChen}, which was grown by a flux method. One single crystal of SmB$_6$ with a large natural surface was cut and polished to a rectangular shape of 1.65 $\times$ 0.60 $\times$ 0.08 mm$^3$. Figure 1(a) is an optical image of the sample. Its large natural surface (1.65 $\times$ 0.60 mm$^2$) was determined to be the (100) plane by x-ray diffraction, as shown in Fig. 1(b) and 1(c). From the x-ray rocking curve in Fig. 1(b), the full width at half maximum (FWHM) is only $0.06^{\circ}$, indicating the high quality of the single crystal. Four silver wires were attached to the sample with silver paint, which were used for both resistivity and thermal conductivity measurements in the (100) plane. The thermal conductivity was measured in a dilution refrigerator, using a standard four-wire steady-state method with two RuO$_2$ chip thermometers, calibrated $in$ $situ$ against a reference RuO$_2$ thermometer. Magnetic fields were applied perpendicular to the (100) plane.

\begin{figure}
\includegraphics[clip,width=9cm]{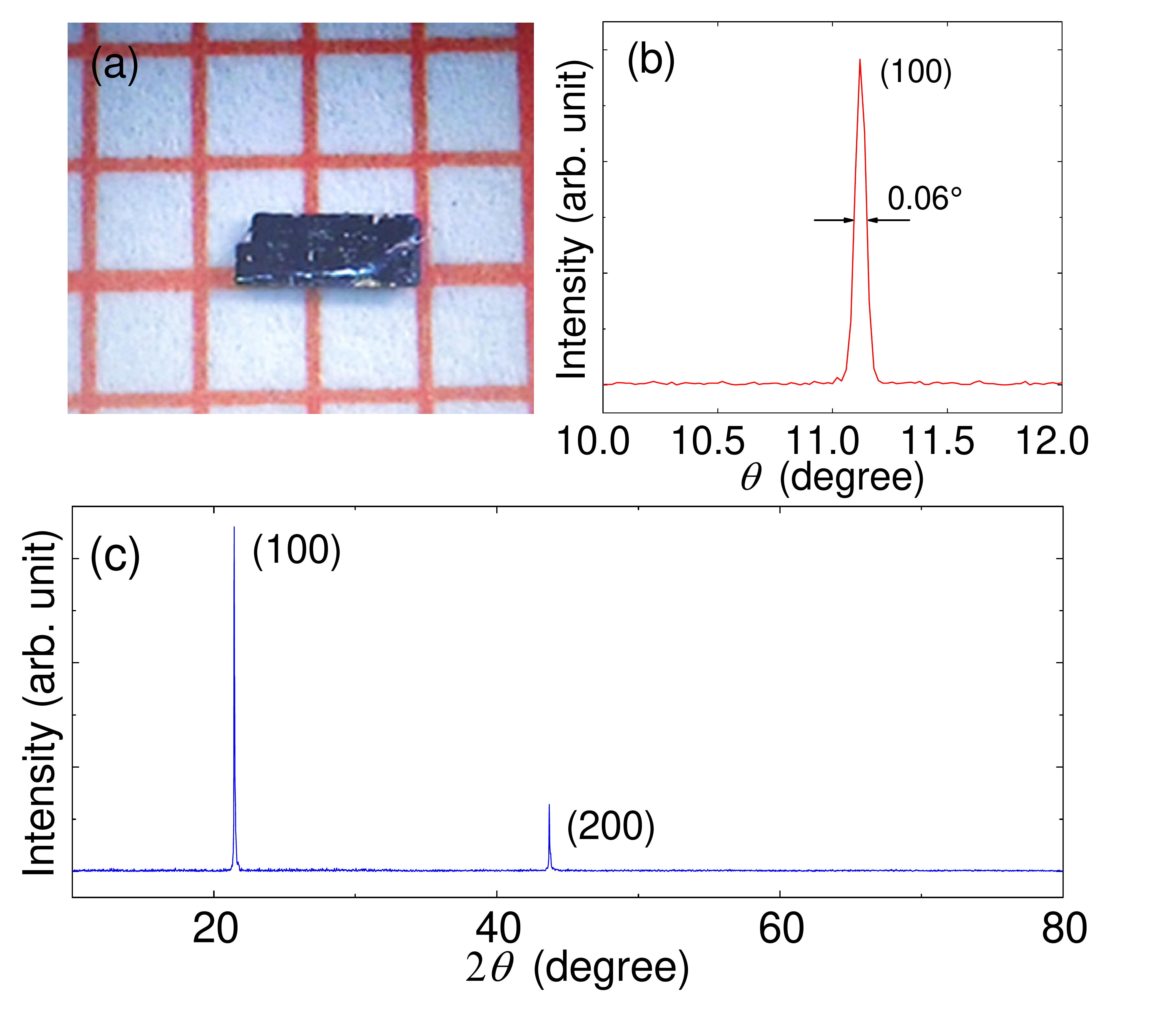}
\caption{(Color online). (a) The optical image of the cut and polished SmB$_6$ single crystal used in this work, with the dimensions of 1.65 $\times$ 0.60 $\times$ 0.08 mm$^3$. Its large natural surface was determined to be the (100) plane by x-ray diffraction, as shown in (b) and (c). (b) The x-ray rocking curve of the (100) Bragg peak. The full width at half maximum (FWHM) is only $0.06^{\circ}$. (c) X-ray diffraction pattern for the large natural surface. Only ($l$00) Bragg peaks show up.
 }
\end{figure}

\begin{figure}
\includegraphics[clip,width=8cm]{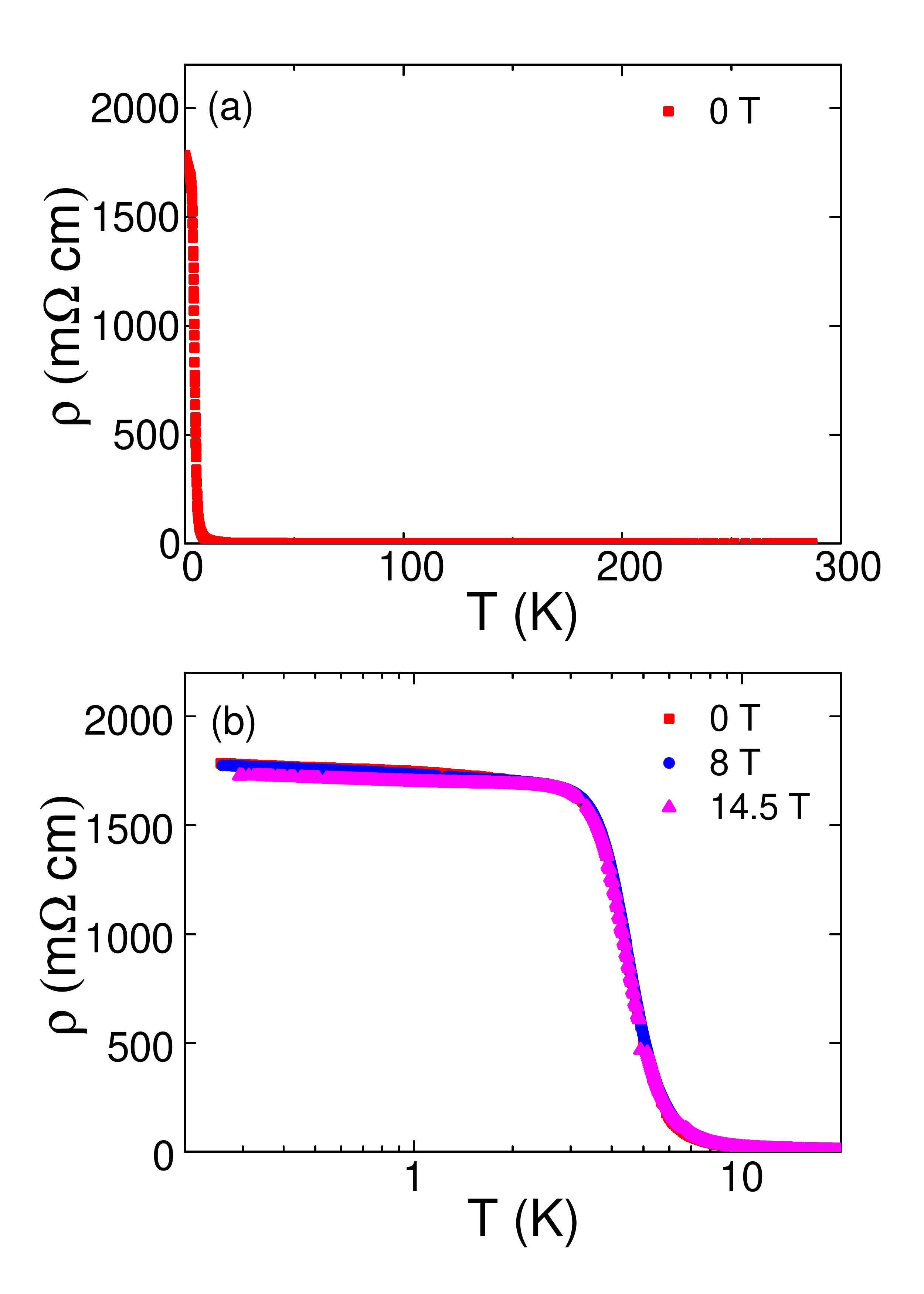}
\caption{(Color online). (a) Temperature dependence of resistivity for the SmB$_6$ single crystal in zero magnetic field. The resistivity increases rapidly with decreasing temperature below 10 K. (b) Low-temperature resistivity in $H$ = 0, 8, and 14.5 T. The resistivity tends to saturate below 3 K, and the effect of magnetic field is very weak in our field range.}
\end{figure}

\begin{figure}
\includegraphics[clip,width=7.77cm]{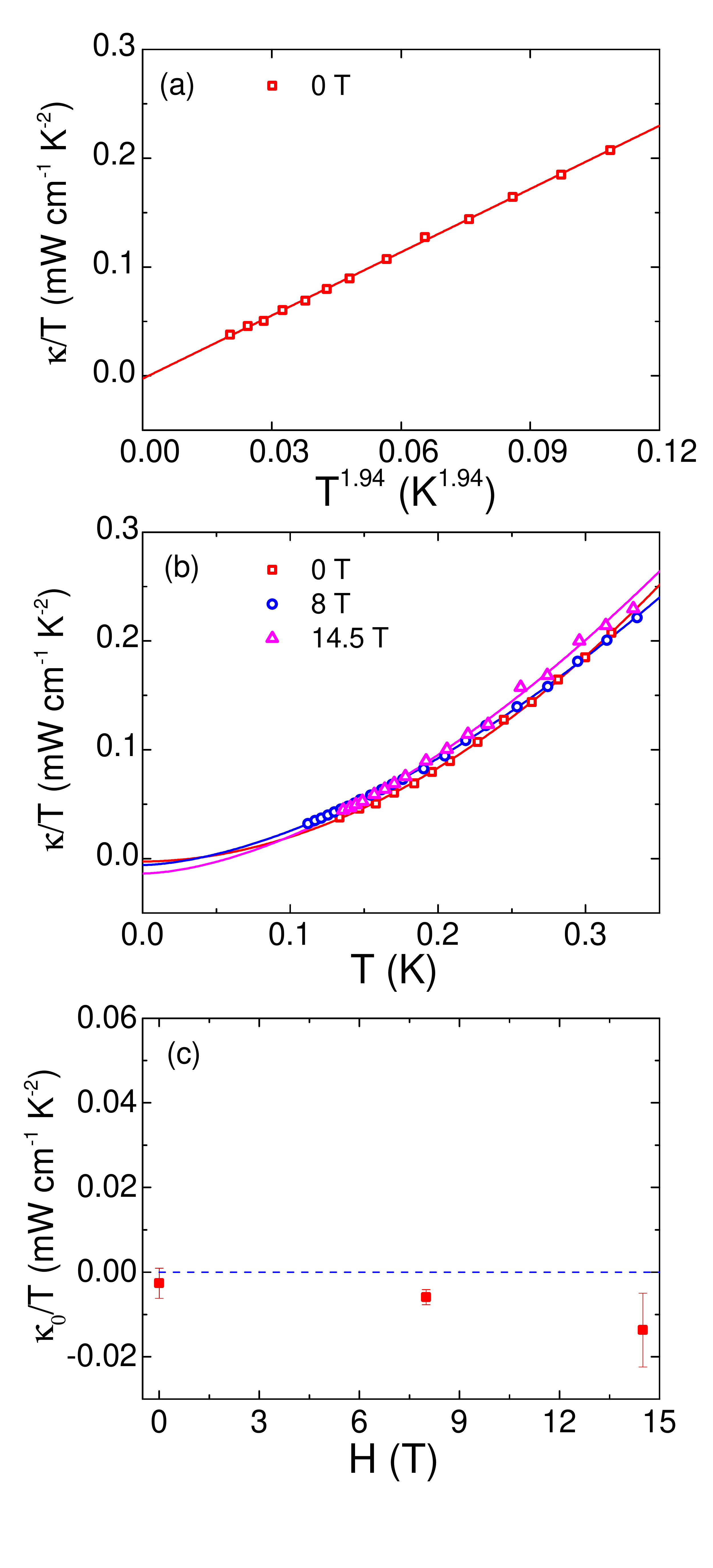}
\caption{(Color online). (a) and (b) Low-temperature thermal conductivity of the SmB$_6$ single crystal in $H$ = 0, 8, and 14.5 T. The solid lines represent the fits to $\kappa/T$ = $a$ + $bT^{\alpha-1}$ below 0.35 K. (c) Field dependence of residual linear term $\kappa_0/T $. It is virtually zero in zero field, and insensitive to magnetic field up to 14.5 T.}
\end{figure}

Figure 2(a) shows the temperature dependence of resistivity $\rho(T)$ for the SmB$_6$ single crystal in zero magnetic field. A sharp increase can be seen below 10 K, as the resistance is dominated by a bulk insulating gap below the Kondo temperature $T_K \simeq$ 50 K \cite{JWAllen}. Below 3 K, a plateau in the $\rho(T)$ curve is observed, as shown in Fig. 2(b). This plateau appears as a consequence of that the resistance of the bulk insulating gap is short circuited by the contribution from the metallic surface states. All these features are consistent with previous measurements on SmB$_6$ \cite{Swolgast,FChen}. The inverse resistivity ratio (IRR) of our sample is $\rho$(0.3K)/$\rho$(290K) = 5115, which is one order of magnitude smaller than that in Refs. \cite{BSTan} and \cite{FChen}. For SmB$_6$, the room-temperature resistivity is dominated by thermal excitations in the insulating bulk, therefore $\rho$(290K) is independent of sample thickness. However, the resistance at very low temperature is dominated by the surface signals, which is nearly independent of the sample thickness, therefore the derived resistivity $\rho$(0.3K) is proportional to the sample thickness. Since our sample is very thin (80 $\mu$m), a much smaller $\rho$(0.3K), thus the IRR, is reasonable. Low-temperature resistivity in magnetic fields $H$ = 0, 8, and 14.5 T are shown in Fig. 2(b). One can see that the effect of magnetic field is very weak in our field range.

The temperature dependence of the thermal conductivity for the SmB$_6$ single crystal in zero field is plotted in Fig. 3(a). The data below 0.35 K are fitted to $\kappa/T$ = $a$ + $bT^{\alpha-1}$, in which the two terms $aT$ and $bT^{\alpha}$ represent contributions from fermionic zero-energy excitations and phonons, respectively \cite{kappa2,kappa}. Because of the specular reflections of phonons at the sample surfaces, the power $\alpha$ in the second term is typically between 2 and 3 \cite{kappa2,kappa}. The fitting gives $\kappa_0/T$ = -0.003 $\pm$ 0.004 mW K$^{-2}$ cm$^{-1}$ and $\alpha$ = 2.94. Comparing with our experimental error bar $\pm$ 0.005 mW K$^{-2}$ cm$^{-1}$, the $\kappa_0/T$ of SmB$_6$ in zero field is virtually zero. Note that for a bulk SmB$_6$ single crystal, the contribution from the metallic surface states to the thermal conductivity is actually negligible. One can estimate this contribution according to the Wiedemann-Franz law, $\kappa/T = L_0/\rho$. By dividing the Lorenz number $L_0$ = 2.45 $\times$ 10$^{-8}$ W$\Omega$K$^{-2}$ with $\rho$(0.3K), we estimate the contribution from the surface states of the order of 0.01 $\mu$W K$^{-2}$ cm$^{-1}$, which is negligible.

The thermal conductivity of the SmB$_6$ single crystal in $H$ = 0, 8, and 14.5 T is plotted in Fig. 3(b). The three curves are almost overlapped with each other. The same fitting process gives $\kappa_0/T =$ -0.006 $\pm$  0.002 mW K$^{-2}$ cm$^{-1}$ and -0.014 $\pm$ 0.009 mW K$^{-2}$ cm$^{-1}$ for $H$ = 8 T and 14.5 T, respectively. The three $\kappa_0/T$ values are plotted in Fig. 3(c).  One can see that the magnetic field barely has any effect on the thermal conductivity of SmB$_6$ up to 14.5 T.

Therefore, the negligible $\kappa_0/T$ in zero and magnetic field excludes the existence of fermionic charge-neutral excitations in bulk SmB$_6$, such as scalar Majorana fermions or spinons. In other words, our thermal conductivity measurements do not support the scenario of a Majorana Fermi liquid state or a spin liquid state in SmB$_6$.

Having settled the debate whether there exist charge-neutral fermionic excitations in bulk SmB$_6$, we turn to discuss other possible scenarios. One possible scenario to reconcile the transport and dHvA data in Ref. \cite{BSTan} is spatial inhomogeneity. In this picture, small metallic domains are formed inside a SmB$_6$ crystal (like islands in the sea), due to either some intrinsic mechanisms (e.g. Kondo breakdown \cite{Alex}) or impurity contributions. Here, because the metallic islands are isolated by the insulating crystal, they can only participate in electrical and thermal transport through the tunneling of electrons, but their magnetization may contribute to the dHvA signals notably.

It was conceived that the reported ``bulk'' dHvA signals may actually come from the metallic surfaces \cite{LiLu}. This possibility is supported by the recent ARPES measurements, which shows that the dHvA signals, reported as from the bulk, coincide almost perfectly with what one should expect from surface states on the (100) and (110) surfaces \cite{LiLu}. This two-dimensional topological surface interpretation is also essentially the basic idea of Ref. \cite{Erten}. This scenario does not directly contradict with the heat-transport results in the present paper.

Two scenarios involving three-dimensional bulk interpretation were proposed recently \cite{Knolle,FaWang}. However, both scenarios suggest the existence of quantum oscillations even for band insulators of certain types: either a simple band insulator of itinerant electrons hybridized with a localized flat band can exhibit quantum oscillations if the cyclotron energy is comparable to the electronic gap \cite{Knolle}; alternatively, an insulator with inverted bands can show quantum oscillations in its bulk low-energy DOS due to the periodic gap narrowing in magnetic fields \cite{FaWang}. Without introducing any bulk excitations or the formation of a bulk Fermi surface (in the scenario of Ref. \cite{FaWang}, the band edges play a similar role as the Fermi surface in metals and the gap periodically narrows, but the gap always remains finite), these scenarios can be compatible with our data.

It was envisaged in Ref. \cite{spinliquid} that an intriguing bulk Fermi liquid (FL) state of conduct electrons with coexisting spin liquids formed by localized moments, the so-called FL$^\ast$, can emerge as the hybridization between the conduction electrons and the local moments vanishes in strong-enough magnetic fields. In this case, there should be a notable $\kappa_0/T$ in the thermal conductivity contributed from conduction electrons, which is not observed here at a magnetic field of 14.5 T. Nonetheless, the quantum oscillations from bulk contribution are already discernible at 14.5 T in the dHvA experiment in Ref. \cite{BSTan}. To verify whether the intriguing FL$^\ast$ state can emerge or not in SmB$_6$ at strong magnetic field, further works of thermal conductivity measurements in higher magnetic fields than 14.5 T are needed.

In summary, we have measured the thermal conductivity of SmB$_6$ single crystal down to 0.1 K. No residual linear term of thermal conductivity was observed at zero field. The thermal conductivity is insensitive to magnetic field up to 14.5 T. These results exclude the existence of fermionic charge-neutral excitations in bulk SmB$_6$, and are undoubtedly helpful for the discrimination between the current theoretical scenarios and the introduction of future ones.

{\it Acknowledgement}: This work is supported in part by the Ministry of Science and Technology of China (National Basic Research Program No: 2012CB821402 and 2015CB921401), the Natural Science Foundation of China, the Program for Professor of Special Appointment (Eastern Scholar) at Shanghai Institutions of Higher Learning, and STCSM of China (No. 15XD1500200). HY was supported by the NSFC under Grant No. 11474175 at Tsinghua University. KS was supported by the National Science Foundation under grants PHY-1402971 at University of Michigan. \\

\noindent $^*$ E-mail: shiyan$\_$li@fudan.edu.cn.

\end{document}